\PassOptionsToPackage{hyphens}{url}
\PassOptionsToPackage{dvipsnames,svgnames,x11names}{xcolor}
\documentclass[
]{article}
\usepackage{xcolor}
\usepackage{amsmath,amssymb}
  \usepackage[T1]{fontenc}
  \usepackage[utf8]{inputenc}
  \usepackage{textcomp} 
\usepackage{lmodern}
\IfFileExists{upquote.sty}{\usepackage{upquote}}{}
\IfFileExists{microtype.sty}{
  \usepackage[]{microtype}
  \UseMicrotypeSet[protrusion]{basicmath} 
}{}
\makeatletter
\@ifundefined{KOMAClassName}{
  \IfFileExists{parskip.sty}{%
    \usepackage{parskip}
  }{
    \setlength{\parindent}{0pt}
    \setlength{\parskip}{6pt plus 2pt minus 1pt}}
}{
  \KOMAoptions{parskip=half}}
\makeatother
\makeatletter
\ifx\paragraph\undefined\else
  \let\oldparagraph\paragraph
  \renewcommand{\paragraph}{
    \@ifstar
      \xxxParagraphStar
      \xxxParagraphNoStar
  }
  \newcommand{\xxxParagraphStar}[1]{\oldparagraph*{#1}\mbox{}}
  \newcommand{\xxxParagraphNoStar}[1]{\oldparagraph{#1}\mbox{}}
\fi
\ifx\subparagraph\undefined\else
  \let\oldsubparagraph\subparagraph
  \renewcommand{\subparagraph}{
    \@ifstar
      \xxxSubParagraphStar
      \xxxSubParagraphNoStar
  }
  \newcommand{\xxxSubParagraphStar}[1]{\oldsubparagraph*{#1}\mbox{}}
  \newcommand{\xxxSubParagraphNoStar}[1]{\oldsubparagraph{#1}\mbox{}}
\fi
\makeatother

\usepackage{color}
\usepackage{fancyvrb}

\DefineVerbatimEnvironment{Highlighting}{Verbatim}{commandchars=\\\{\}}
\usepackage{framed}
\definecolor{shadecolor}{RGB}{241,243,245}
\newenvironment{Shaded}{\begin{snugshade}}{\end{snugshade}}

\newcommand{\CommentTok}[1]{\textcolor[rgb]{0.37,0.37,0.37}{#1}}

\newcommand{\DataTypeTok}[1]{\textcolor[rgb]{0.68,0.00,0.00}{#1}}
\newcommand{\DecValTok}[1]{\textcolor[rgb]{0.68,0.00,0.00}{#1}}

\newcommand{\ErrorTok}[1]{\textcolor[rgb]{0.68,0.00,0.00}{#1}}
\newcommand{\ExtensionTok}[1]{\textcolor[rgb]{0.00,0.23,0.31}{#1}}
\newcommand{\FloatTok}[1]{\textcolor[rgb]{0.68,0.00,0.00}{#1}}
\newcommand{\FunctionTok}[1]{\textcolor[rgb]{0.28,0.35,0.67}{#1}}

\newcommand{\NormalTok}[1]{\textcolor[rgb]{0.00,0.23,0.31}{#1}}

\newcommand{\OtherTok}[1]{\textcolor[rgb]{0.00,0.23,0.31}{#1}}

\newcommand{\StringTok}[1]{\textcolor[rgb]{0.13,0.47,0.30}{#1}}

\usepackage{longtable,booktabs,array}
\usepackage{calc} 
\usepackage{etoolbox}
\makeatletter
\patchcmd\longtable{\par}{\if@noskipsec\mbox{}\fi\par}{}{}
\makeatother
\IfFileExists{footnotehyper.sty}{\usepackage{footnotehyper}}{\usepackage{footnote}}
\makesavenoteenv{longtable}
\usepackage{graphicx}
\makeatletter
\newsavebox\pandoc@box
\newcommand*\pandocbounded[1]{
  \sbox\pandoc@box{#1}%
  \Gscale@div\@tempa{\textheight}{\dimexpr\ht\pandoc@box+\dp\pandoc@box\relax}%
  \Gscale@div\@tempb{\linewidth}{\wd\pandoc@box}%
  \ifdim\@tempb\p@<\@tempa\p@\let\@tempa\@tempb\fi
  \ifdim\@tempa\p@<\p@\scalebox{\@tempa}{\usebox\pandoc@box}%
  \else\usebox{\pandoc@box}%
  \fi%
}
\def\fps@figure{htbp}
\makeatother

\providecommand{\tightlist}{%
  \setlength{\itemsep}{0pt}\setlength{\parskip}{0pt}}

\usepackage{arxiv}
\usepackage{orcidlink}
\usepackage{amsmath}
\usepackage[T1]{fontenc}
\makeatletter
\@ifpackageloaded{caption}{}{\usepackage{caption}}
\AtBeginDocument{%
\ifdefined\contentsname
  \renewcommand*\contentsname{Table of contents}
\else
  \newcommand\contentsname{Table of contents}
\fi
\ifdefined\listfigurename
  \renewcommand*\listfigurename{List of Figures}
\else
  \newcommand\listfigurename{List of Figures}
\fi
\ifdefined\listtablename
  \renewcommand*\listtablename{List of Tables}
\else
  \newcommand\listtablename{List of Tables}
\fi
\ifdefined\figurename
  \renewcommand*\figurename{Figure}
\else
  \newcommand\figurename{Figure}
\fi
\ifdefined\tablename
  \renewcommand*\tablename{Table}
\else
  \newcommand\tablename{Table}
\fi
}
\@ifpackageloaded{float}{}{\usepackage{float}}
\floatstyle{ruled}
\@ifundefined{c@chapter}{\newfloat{codelisting}{h}{lop}}{\newfloat{codelisting}{h}{lop}[chapter]}
\floatname{codelisting}{Listing}

\makeatother
\makeatletter
\@ifpackageloaded{caption}{}{\usepackage{caption}}
\@ifpackageloaded{subcaption}{}{\usepackage{subcaption}}
\makeatother
\usepackage{bookmark}
\IfFileExists{xurl.sty}{\usepackage{xurl}}{} 
\hypersetup{
  pdftitle={AI-Reporter: A Path to a New Genre of Scientific Communication},
  pdfauthor={Gerd Gra\ss hoff},
  pdfkeywords={scientific communication, artificial
intelligence, automated publishing, large language models, academic
presentations, knowledge preservation},
  colorlinks=true,
  linkcolor={blue},
  filecolor={Maroon},
  citecolor={Blue},
  urlcolor={Blue},
  pdfcreator={LaTeX via pandoc}}

\newcommand{\runninghead}{A Preprint }
\renewcommand{\runninghead}{AI-Reporter Pipeline }
\title{AI-Reporter: A Path to a New Genre of Scientific Communication}
\usepackage{etoolbox}
\makeatletter
\providecommand{\subtitle}[1]{
  \apptocmd{\@title}{\par {\large #1 \par}}{}{}
}
\makeatother
\subtitle{From Presentation to Publication through Automation
Intelligence}
\author{\textbf{Gerd
Gra\ss hoff}~\orcidlink{0000-0002-4450-357X}\\Department of
Philosophy\\Humboldt-Universität zu
Berlin\\Berlin,\ 10099\\\href{mailto:gerd.grasshoff@hu-berlin.de}{gerd.grasshoff@hu-berlin.de}}
\date{}
\begin{document}
\maketitle
\begin{abstract}
The AI-Reporter represents a paradigmatic shift in scientific
publication practice. This document demonstrates through a concrete case
study how our system transforms academic presentations into
publication-ready chapters---in less than three minutes. Using Arno
Simons' lecture on Large Language Models (LLMs) from the ``Large
Language Models for the History, Philosophy, and Sociology of Science''
workshop as an example, we show how technological innovation bridges the
gap between ephemeral presentation and permanent scientific
documentation.
\end{abstract}
{\bfseries \emph Keywords}
\def\sep{\textbullet\ }
scientific communication \sep artificial intelligence \sep automated
publishing \sep large language models \sep academic presentations \sep 
knowledge preservation

\section{The Vision: A New Genre of Scientific
Communication}\label{the-vision-a-new-genre-of-scientific-communication}

In the academic world, there exists a fundamental asymmetry: While
preparing a scientific presentation takes weeks or months, the insights
presented often vanish into the ether immediately after the talk. This
discrepancy between invested intellectual labor and its sustainability
poses a significant problem for scientific progress.

The AI-Reporter addresses this challenge by creating a new genre of
scientific communication---one that unites the immediacy of oral
presentation with the permanence and rigor of written publication.

\subsection{The Goal: Beyond Preservation---Expanding Scientific
Communication}\label{the-goal-beyond-preservationexpanding-scientific-communication}

Our vision extends far beyond merely preserving presentations. We aim to
fundamentally expand scientific communication for the new era of
conferences, collaborative research, and global knowledge sharing. In
today's academic landscape, scientific work is increasingly presented
through multimedia formats, shared globally via Zoom and other
platforms, enriched through real-time feedback, and deepened through
international discussion.

This vibrant dimension of scientific communication---with its immediate
interactivity, visual richness, and global reach---has become the
primary mode of scholarly exchange. Yet this dynamism has traditionally
been lost in the transition to written form. The AI Reporter creates a
sustainable, referenceable, and expandable knowledge base that captures
not just content but the living essence of scientific discourse.

Our goal encompasses:

\begin{itemize}
\tightlist
\item
  \textbf{Comprehensive Model of Scientific Communication}: Integration
  of oral, visual, and multimedia information channels into a textually
  robust, scientifically verifiable, and sustainable form of scholarly
  communication
\item
  \textbf{Integral Transformation}: Through AI-supported reformulation,
  the complete message and information of the presentation---including
  its temporal rhythm and dynamic flow---is transformed into text that
  maintains the vitality of the oral delivery while ensuring scientific
  accuracy and completeness for future readers
\item
  \textbf{Practical Efficiency}: Transformation occurs in minutes rather
  than weeks
\item
  \textbf{Scientific Standards}: The result meets the quality
  requirements of academic publications and peer review
\item
  \textbf{Global Accessibility}: From presentation to worldwide open
  publication with minimal delay
\item
  \textbf{Living Documentation}: Creating a referenceable, citable, and
  expandable foundation for future research
\end{itemize}

The deployment of advanced AI tools aims to accelerate the processing
pipeline to such a degree that the journey from presentation to global
publication requires only the author's consent. With adapted technical
processing channels and modules, we envision a future where:

\begin{itemize}
\tightlist
\item
  A morning conference presentation becomes an afternoon publication
\item
  Live Zoom seminars transform into accessible chapters within hours
\item
  Preliminary findings shared in workshops become citable references
  immediately
\item
  The entire scientific community gains instant access to cutting-edge
  research
\end{itemize}

This represents not just an optimization of existing processes, but a
fundamental reimagining of how scientific knowledge is created, shared,
and preserved in the 21st century.

\subsection{The Methodology: A New
Approach}\label{the-methodology-a-new-approach}

Our methodological innovation consists in reversing the traditional
transcription process. Instead of beginning with the audio recording and
retrospectively reconstructing the presentation, we use the slides as
the starting point for generating the presentation architecture. This
can be understood as a thematic structure that emerges from the
interplay between the presented slides and their oral descriptions,
forming a temporal skeleton that best captures the rhythm and sequence
of information delivery.

This approach is based on the insight that academic presentations follow
a dual logic: The visual level structures and anchors the argumentation,
while the verbal level unfolds and contextualizes it. Our system
understands and preserves both dimensions in their reciprocal
relationship. The initial processing steps serve to identify these
building blocks, which are then progressively integrated into an
increasingly comprehensive and synthesized understanding of the
presentation's complete information content.

\section{Case Discussion: Opening Presentation by Arno
Simons}\label{case-discussion-opening-presentation-by-arno-simons}

To demonstrate the functionality of our system, we selected the opening
presentation by Arno Simons on ``Large Language Models for the History,
Philosophy, and Sociology of Science'' from the NEPI workshop at TU
Berlin. We are grateful to Arno Simons for his permission to use his
contribution for this detailed documentation of the processing workflow.
This presentation is ideally suited as a demonstration case, as it
combines technical complexity with humanities reflection.

\begin{figure}

\centering{

\includegraphics[width=0.6\linewidth,height=\textheight,keepaspectratio]{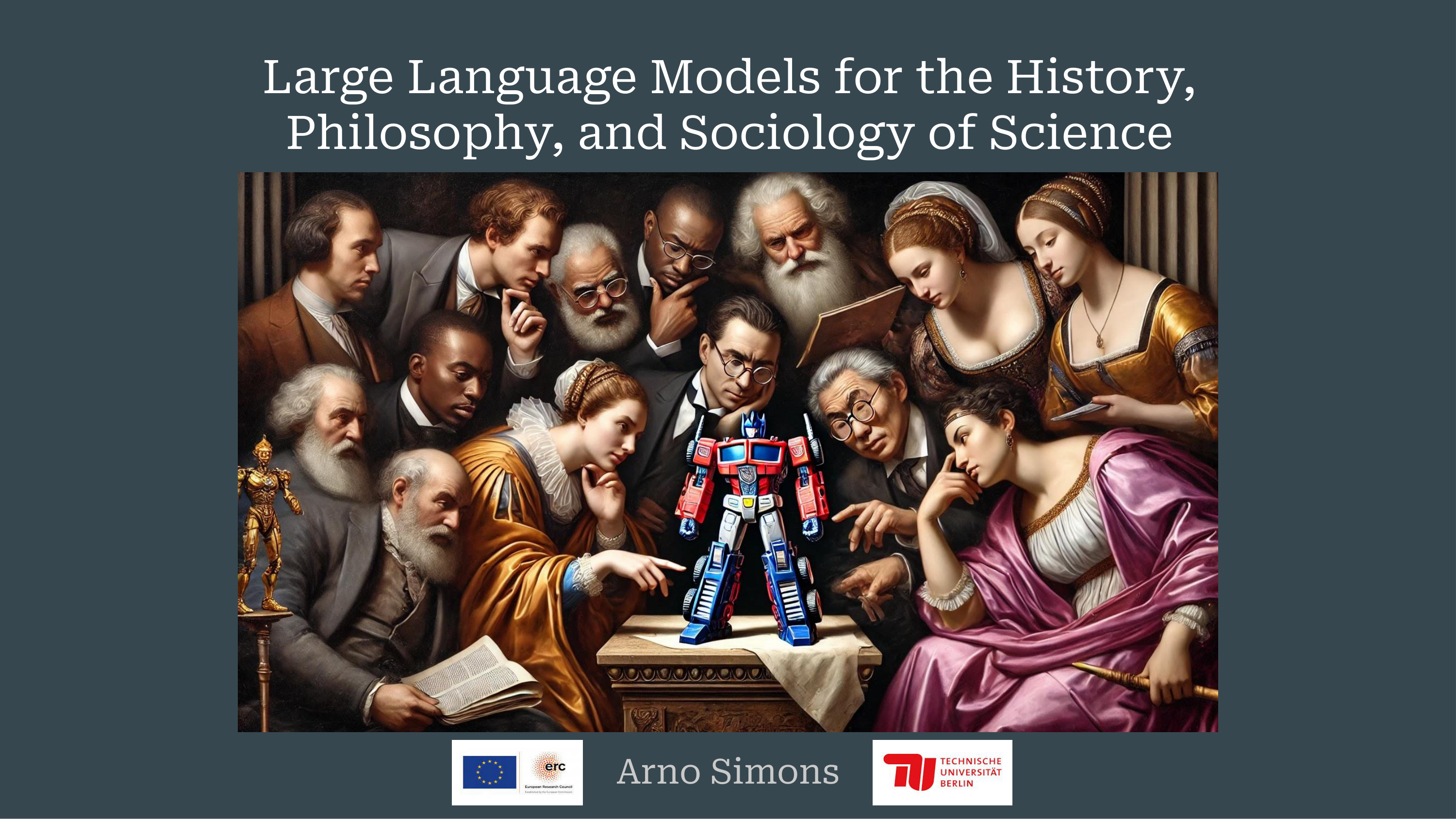}

}

\caption{\label{fig-title}The title slide shows the connection between
traditional humanities and modern AI technology}

\end{figure}%

The visual metaphor of the title slide---historical scholars sitting
together with a Transformer robot at a table---already anticipates the
central thesis of the lecture: the productive synthesis of centuries-old
humanistic tradition and cutting-edge technology.

\section{The Processing Workflow: A Systematic
Analysis}\label{the-processing-workflow-a-systematic-analysis}

\subsection{Step 0: Validation of Source
Materials}\label{step-0-validation-of-source-materials}

Every scientific method begins with careful examination of its
prerequisites. Our system implements a rigorous validation phase:

\begin{Shaded}
\begin{Highlighting}[]
\FunctionTok{\{}
  \DataTypeTok{"presentation\_id"}\FunctionTok{:} \StringTok{"003"}\FunctionTok{,}
  \DataTypeTok{"validation\_results"}\FunctionTok{:} \FunctionTok{\{}
    \DataTypeTok{"pdf\_integrity"}\FunctionTok{:} \StringTok{"verified"}\FunctionTok{,}
    \DataTypeTok{"video\_quality"}\FunctionTok{:} \StringTok{"sufficient"}\FunctionTok{,}
    \DataTypeTok{"metadata\_completeness"}\FunctionTok{:} \StringTok{"confirmed"}\FunctionTok{,}
    \DataTypeTok{"author\_information"}\FunctionTok{:} \StringTok{"complete"}
  \FunctionTok{\},}
  \DataTypeTok{"processing\_status"}\FunctionTok{:} \StringTok{"READY"}
\FunctionTok{\}}
\end{Highlighting}
\end{Shaded}

This seemingly trivial verification prevents costly failed attempts and
ensures the quality of the final product.

\subsection{Step 1: Extraction of the Visual Knowledge
Architecture}\label{step-1-extraction-of-the-visual-knowledge-architecture}

The extraction of presentation slides occurs with highest precision---17
image files at 300 DPI resolution. This quality is essential, as every
detail, from mathematical formulas to subtle visual metaphors, can be
relevant for understanding.

The prioritization of the visual level reflects an epistemological
insight: In academic presentations, slides function not as mere
illustrations, but as independent knowledge carriers that represent
complex relationships in condensed form.

\subsection{Step 2: Temporal Mapping}\label{step-2-temporal-mapping}

The temporal assignment of slides to the video timeline requires precise
frame-by-frame analysis. Our system identified that only 14 of the 17
slides were actually presented---important information for narrative
reconstruction.

\begin{longtable}[]{@{}llll@{}}
\toprule\noalign{}
Slide & Content & Timestamp & Duration \\
\midrule\noalign{}
\endhead
\bottomrule\noalign{}
\endlastfoot
01 & Title & 00:00 & 8 seconds \\
02 & Agenda & 00:08 & 33 seconds \\
03 & Transformer Architecture & 00:41 & 1:55 minutes \\
\ldots{} & \ldots{} & \ldots{} & \ldots{} \\
\end{longtable}

This temporal mapping enables precise synchronization of visual and
verbal information.

\subsection{Step 3: Semantic Analysis of Slide
Content}\label{step-3-semantic-analysis-of-slide-content}

The actual innovation lies in our Partitur system (developed by Gerd
Gra\ss hoff, OpenScienceTechnology), which orchestrates complex processing
pipelines. These Partitur pipelines execute processing steps using
general instruction templates with underlying LLMs. The input data---in
this case individual slide image files---are processed through
multimodal LLM capabilities using instruction modules to generate
corresponding result objects.

Let us consider the Transformer architecture slide:

\begin{figure}

\centering{

\includegraphics[width=0.6\linewidth,height=\textheight,keepaspectratio]{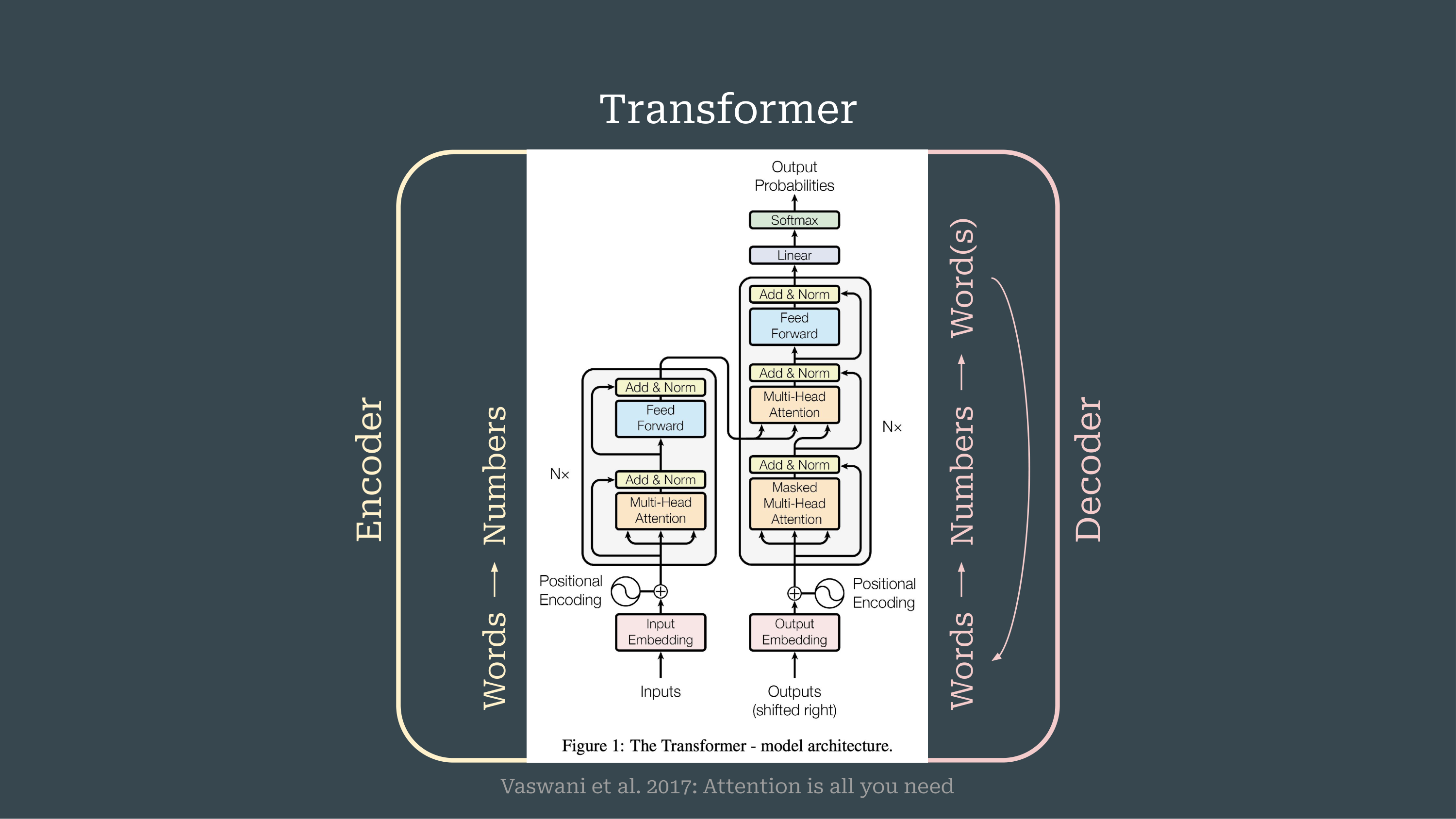}

}

\caption{\label{fig-transformer}The Transformer architecture as the
foundation of modern language models}

\end{figure}%

The Partitur system processes this slide through a multi-stage pipeline:

\begin{figure}

\centering{

\pandocbounded{\includegraphics[keepaspectratio]{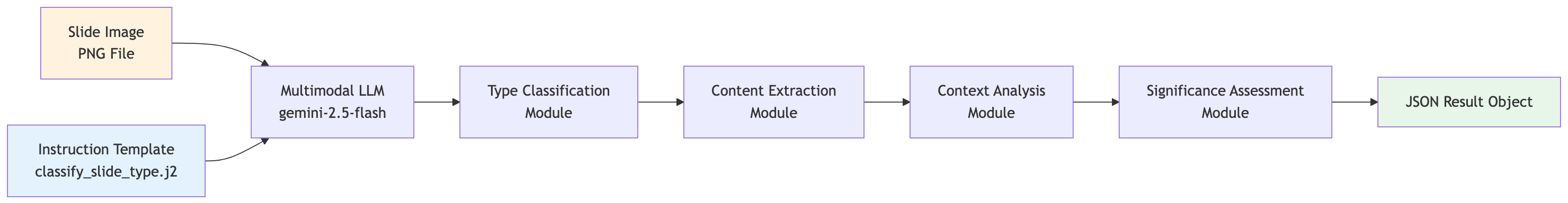}}

}

\caption{\label{fig-partitur-pipeline}Partitur pipeline for slide
analysis}

\end{figure}%

Each pipeline component contributes specific analysis:

\begin{enumerate}
\def\labelenumi{\arabic{enumi}.}
\tightlist
\item
  \textbf{Input Processing}: The PNG image file is loaded with full
  resolution
\item
  \textbf{Instruction Integration}: The classify\_slide\_type.j2
  template provides structured prompts
\item
  \textbf{Type Classification}: Identifies slide category
  (technical\_architecture, conceptual, data, etc.)
\item
  \textbf{Content Extraction}: Summarizes visual and textual elements
\item
  \textbf{Context Analysis}: Interprets significance within presentation
  narrative
\item
  \textbf{Significance Assessment}: Evaluates academic importance
\end{enumerate}

This pipeline generates the following result object:

\begin{Shaded}
\begin{Highlighting}[]
\FunctionTok{\{}
  \DataTypeTok{"slide\_type"}\FunctionTok{:} \StringTok{"technical\_architecture"}\FunctionTok{,}
  \DataTypeTok{"content\_summary"}\FunctionTok{:} \StringTok{"Transformer model with encoder{-}decoder structure"}\FunctionTok{,}
  \DataTypeTok{"comprehensive\_analysis"}\FunctionTok{:} \StringTok{"This slide presents the groundbreaking }
\StringTok{    Transformer architecture by Vaswani et al. (2017). The diagram }
\StringTok{    illustrates bidirectional processing in the encoder and }
\StringTok{    sequential generation in the decoder. The multi{-}head attention }
\StringTok{    mechanisms enable parallel processing of contextual }
\StringTok{    information—a fundamental innovation for understanding }
\StringTok{    natural language."}\FunctionTok{,}
  \DataTypeTok{"academic\_significance"}\FunctionTok{:} \StringTok{"high"}
\FunctionTok{\}}
\end{Highlighting}
\end{Shaded}

\subsection{Step 4: Intelligent Slide Curation - The Art of Academic
Transformation}\label{step-4-intelligent-slide-curation---the-art-of-academic-transformation}

The curation phase represents a critical innovation in our pipeline,
addressing one of the fundamental challenges in transforming
presentations into publications. Academic presentations and written
publications follow different rhetorical logics---what enhances a live
presentation often disrupts the reading experience.

\subsubsection{The Curation Challenge}\label{the-curation-challenge}

Presentations employ various techniques that serve oral delivery but
complicate written documentation:

\begin{itemize}
\tightlist
\item
  \textbf{Progressive Reveal}: Speakers build complex diagrams over
  multiple slides
\item
  \textbf{Section Dividers}: ``Part 2: Methods'' slides that segment
  oral presentation
\item
  \textbf{Interactive Elements}: ``Questions?'' or ``Let's discuss''
  slides
\item
  \textbf{Redundant Frames}: Title slides, thank you slides, contact
  information
\end{itemize}

Our Partitur system implements sophisticated filtering algorithms that
automatically identify and handle these presentation-specific elements:

\begin{figure}

\centering{

\pandocbounded{\includegraphics[keepaspectratio]{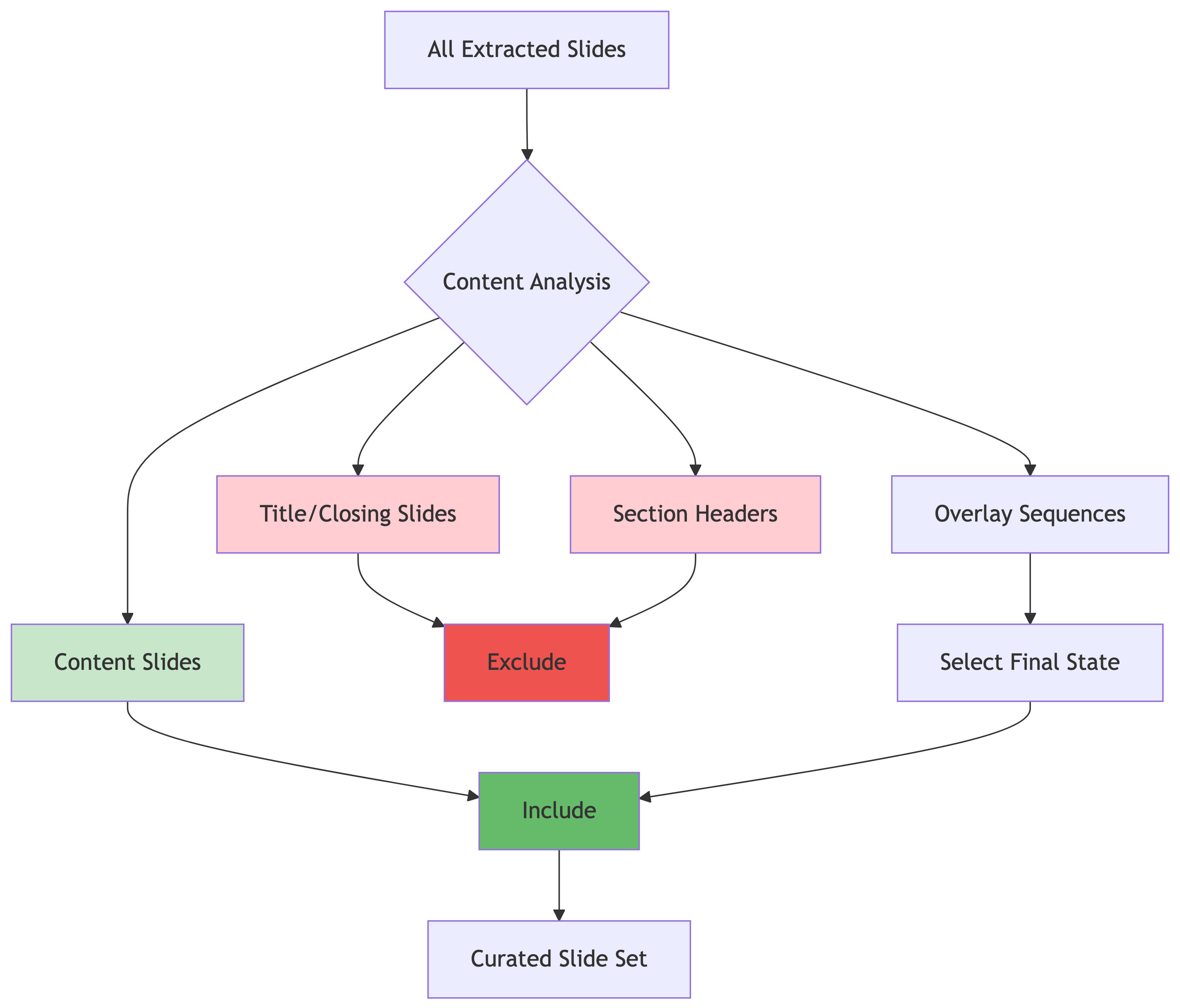}}

}

\caption{\label{fig-curation}Intelligent slide curation process}

\end{figure}%

\subsubsection{Overlay Handling
Excellence}\label{overlay-handling-excellence}

Consider a typical progressive reveal sequence where a complex diagram
is built over 5 slides: - Slide 14: Base diagram - Slide 15: + First
annotation - Slide 16: + Second annotation\\
- Slide 17: + Third annotation - Slide 18: Complete diagram with all
elements

Our system intelligently selects only Slide 18 for the publication,
preserving the complete information while eliminating redundancy. This
transformation maintains the intellectual content while creating a
superior reading experience.

\subsubsection{Manual Control and
Flexibility}\label{manual-control-and-flexibility}

While the automatic curation achieves excellent results, the system
preserves complete editorial control through the
\texttt{slides\_for\_production.json} file. Researchers can: - Override
automatic decisions - Include specific transition slides if academically
relevant - Exclude content slides that may be too context-specific -
Fine-tune the selection for their particular discipline's conventions

This balance between intelligent automation and human expertise ensures
that each publication meets the highest academic standards while
respecting disciplinary differences.

\subsection{Step 5: Precise Slide-Video
Synchronization}\label{step-5-precise-slide-video-synchronization}

Before transcription can begin, our system must solve a fundamental
challenge: establishing perfect temporal alignment between the visual
presentation and the spoken word. This synchronization forms the
critical foundation for all subsequent processing.

\subsubsection{The Temporal Alignment
Challenge}\label{the-temporal-alignment-challenge}

In live presentations, speakers rarely announce slide transitions
explicitly. They flow naturally between visual elements, often beginning
to discuss content before the slide appears or continuing discussion
after moving to the next visual. Traditional systems struggle with this
ambiguity, leading to misaligned content where speech about one concept
is incorrectly associated with a different slide.

Our multi-stage synchronization process achieves precision through
intelligent visual matching:

\begin{figure}

\centering{

\pandocbounded{\includegraphics[keepaspectratio]{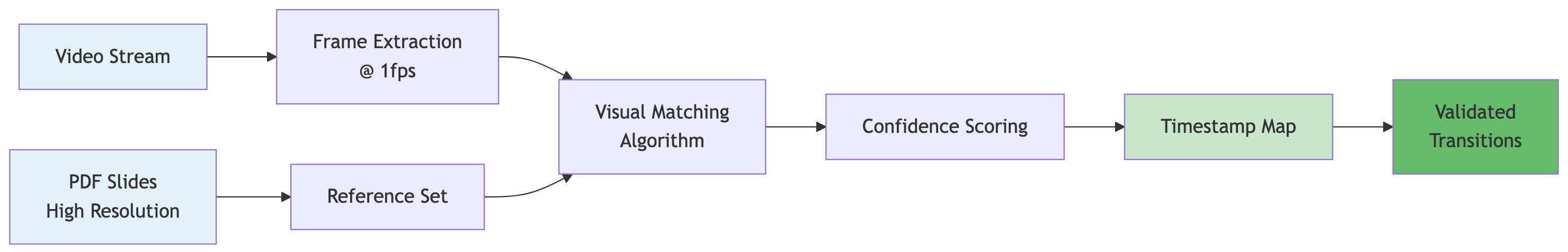}}

}

\caption{\label{fig-sync}Precision synchronization pipeline}

\end{figure}%

The system processes thousands of video frames, comparing each against
the reference slide set to detect exact transition moments. This creates
a temporal map with millisecond precision:

\begin{Shaded}
\begin{Highlighting}[]
\FunctionTok{\{}
  \DataTypeTok{"slide\_transitions"}\FunctionTok{:} \FunctionTok{\{}
    \DataTypeTok{"slide\_01"}\FunctionTok{:} \FunctionTok{\{}
      \DataTypeTok{"timestamp"}\FunctionTok{:} \StringTok{"00:00:00"}\FunctionTok{,}
      \DataTypeTok{"confidence"}\FunctionTok{:} \FloatTok{0.98}\FunctionTok{,}
      \DataTypeTok{"duration\_until\_next"}\FunctionTok{:} \StringTok{"00:00:08"}
    \FunctionTok{\},}
    \DataTypeTok{"slide\_03"}\FunctionTok{:} \FunctionTok{\{}
      \DataTypeTok{"timestamp"}\FunctionTok{:} \StringTok{"00:00:41"}\FunctionTok{,}
      \DataTypeTok{"confidence"}\FunctionTok{:} \FloatTok{0.95}\FunctionTok{,}
      \DataTypeTok{"duration\_until\_next"}\FunctionTok{:} \StringTok{"00:01:55"}
    \FunctionTok{\}}
  \FunctionTok{\}}
\FunctionTok{\}}
\end{Highlighting}
\end{Shaded}

This precision ensures that every spoken word is correctly associated
with its visual context, enabling the system to understand not just what
was said, but what was being shown when it was said.

\subsection{Step 6: Transcription with Contextual
Intelligence}\label{step-6-transcription-with-contextual-intelligence}

With perfect synchronization established, transcription goes beyond
mechanical reproduction of the spoken word. It preserves academic
precision while transforming oral spontaneity into written eloquence:

\begin{verbatim}
Original audio: "So, um, this is the famous, uh, Transformer architecture. 
You may know the Transformer. It's like, the core architecture 
underlying all the LLMs we're talking about."

Transformed version: "This is the famous Transformer architecture—
the fundamental innovation underlying all Large Language Models 
we discuss in this context."
\end{verbatim}

\subsection{Step 7: Narrative
Synthesis}\label{step-7-narrative-synthesis}

The integration of visual and verbal levels represents a crucial
transformation point in our pipeline. Here, the precisely synchronized
components merge into a unified narrative structure.

\subsubsection{The Storyboarding
Architecture}\label{the-storyboarding-architecture}

Our storyboarding system creates what we call ``narrative
blocks''---coherent units that combine: - The visual anchor (slide) -
The temporal context (when it appeared) - The verbal elaboration (what
was said about it)

\begin{figure}

\centering{

\pandocbounded{\includegraphics[keepaspectratio]{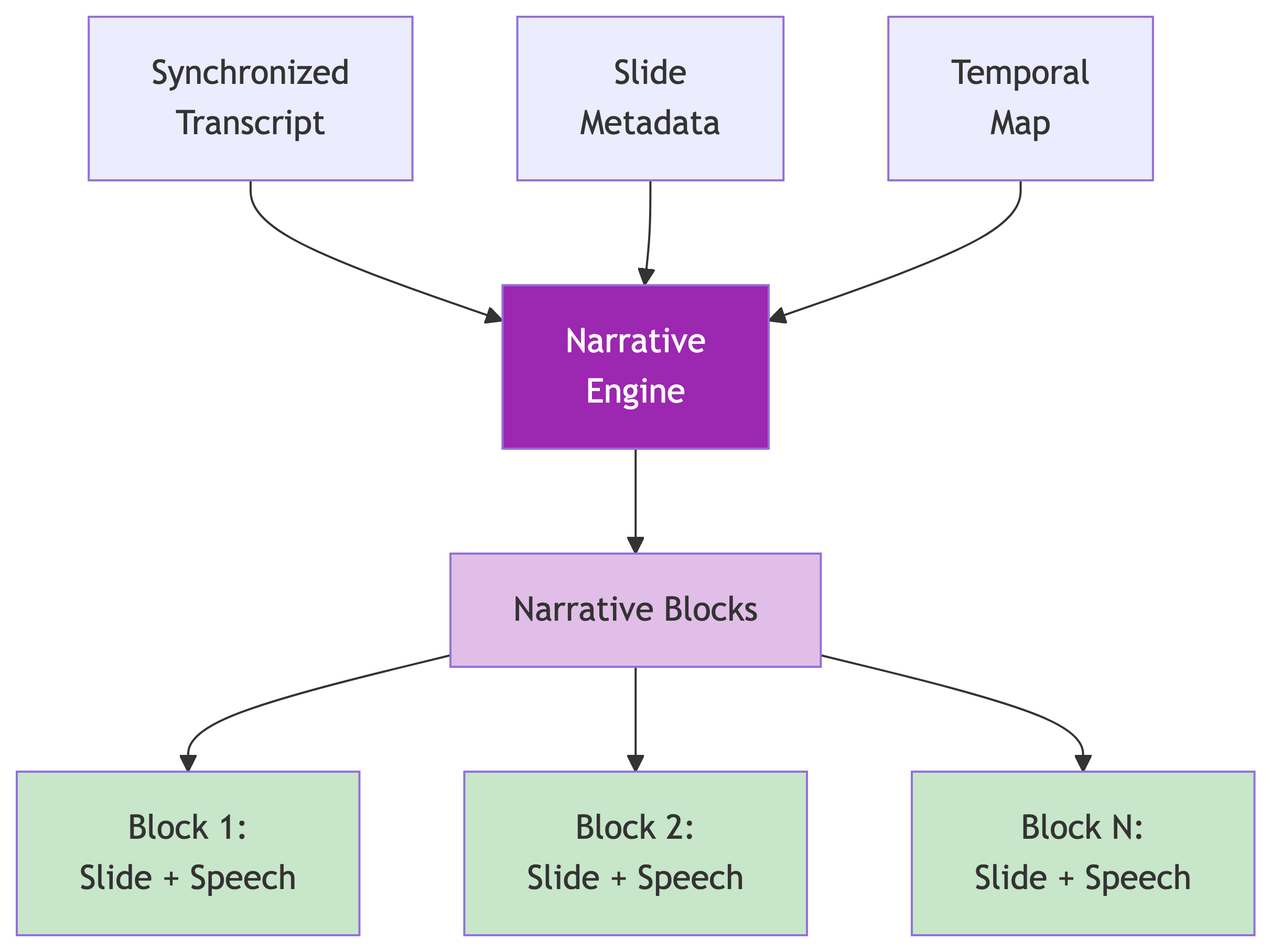}}

}

\caption{\label{fig-storyboard}Narrative synthesis through
storyboarding}

\end{figure}%

This architecture ensures that the final narrative maintains the
presenter's intended flow while adapting to the requirements of written
academic discourse. Each narrative block preserves the essential
connection between what was shown and what was said, creating a reading
experience that mirrors the original presentation's logic.

\subsection{Step 8: Academic Contextualization - From Temporal Narrative
to Thematic
Synthesis}\label{step-8-academic-contextualization---from-temporal-narrative-to-thematic-synthesis}

The transformation from storyboard to content report represents the most
intellectually sophisticated step in our pipeline. Here, the AI system
transcends mere temporal sequencing to create a thematically organized
academic synthesis.

\subsubsection{The Synthesis Process: ID 003 as
Exemplar}\label{the-synthesis-process-id-003-as-exemplar}

The transformation from storyboard to content report represents a
sophisticated reorganization of knowledge. Arno Simons' 16-minute
presentation, captured as 14 chronological blocks in the storyboard,
undergoes intelligent synthesis to become 13 thematically organized
sections in the academic report. This is not a simple one-to-one mapping
but a complex transformation that:

\begin{itemize}
\tightlist
\item
  \textbf{Merges related content} from different time points into
  coherent themes
\item
  \textbf{Generates new content} like the overview that synthesizes the
  entire presentation
\item
  \textbf{Reorganizes information} from temporal sequence to logical
  academic structure
\item
  \textbf{Elevates language} from spoken informality to written
  scholarly prose
\end{itemize}

\textbf{Figure: Transformation from chronological presentation flow to
thematic academic structure}

\begin{verbatim}
ORIGINAL PRESENTATION                    SYNTHESIZED REPORT
(14 Temporal Blocks)                     (13 Thematic Sections)

1. Opening remarks        -----+         
                              |---->     1. Overview synthesis
2. Agenda slide          -----+              (from entire talk)

3. Transformer architecture ===+
                              |====>     2. Foundational Architecture
4. Encoder explanation     ===+              (combines blocks 3-4)
                              |
                              |---->     3-6. Model Types Analysis
5. BERT model             ----|              (reorganizes blocks 4-6)
                              |
6. GPT model              ----+

7-9. Model evolution      ====+====>     7. Scientific LLM Evolution
                              |
                              +====>     8. Adaptation Strategies

10-11. Survey & apps      =========>     10. HPSS Survey Findings

12. RAG systems           =========>     9. RAG Pipeline Systems

13-14. Challenges         =========>     11-13. Future Directions

Legend: - = contributing content  = = primary transformation
\end{verbatim}

The diagram above shows the transformation patterns. Here are the
specific details:

\begin{longtable}[]{@{}
  >{\raggedright\arraybackslash}p{(\linewidth - 8\tabcolsep) * \real{0.3158}}
  >{\raggedright\arraybackslash}p{(\linewidth - 8\tabcolsep) * \real{0.1184}}
  >{\raggedright\arraybackslash}p{(\linewidth - 8\tabcolsep) * \real{0.0395}}
  >{\raggedright\arraybackslash}p{(\linewidth - 8\tabcolsep) * \real{0.2763}}
  >{\raggedright\arraybackslash}p{(\linewidth - 8\tabcolsep) * \real{0.2500}}@{}}
\toprule\noalign{}
\begin{minipage}[b]{\linewidth}\raggedright
Original Blocks (Time)
\end{minipage} & \begin{minipage}[b]{\linewidth}\raggedright
Content
\end{minipage} & \begin{minipage}[b]{\linewidth}\raggedright
→
\end{minipage} & \begin{minipage}[b]{\linewidth}\raggedright
Synthesized Sections
\end{minipage} & \begin{minipage}[b]{\linewidth}\raggedright
Transformation Type
\end{minipage} \\
\midrule\noalign{}
\endhead
\bottomrule\noalign{}
\endlastfoot
Block 1-2 (00:00-00:41) & Opening \& Agenda & → & Section 1 & Complete
synthesis from entire presentation \\
Block 3-4 (00:41-03:07) & Transformer \& Encoder & → & Section 2 &
Combined and enriched with context \\
Block 4-6 (02:36-05:08) & Encoder, BERT, GPT & → & Sections 3-6 &
Reorganized by model type \\
Block 7-9 (05:08-09:50) & Model evolution & → & Sections 7-8 & Split
into evolution \& adaptation \\
Block 10-11 (09:50-11:21) & Survey \& applications & → & Section 10 &
Synthesized findings \\
Block 12 (11:21-12:38) & RAG systems & → & Section 9 & Enriched with
technical detail \\
Block 13-14 (12:38-15:53) & Challenges & → & Sections 11-13 & Thematic
expansion \\
\end{longtable}

\paragraph{From Storyboard: Raw Temporal
Sequence}\label{from-storyboard-raw-temporal-sequence}

The storyboard preserves the exact flow of the presentation:

\begin{Shaded}
\begin{Highlighting}[]
\FunctionTok{\{}
  \DataTypeTok{"block"}\FunctionTok{:} \DecValTok{3}\FunctionTok{,}
  \DataTypeTok{"slide"}\FunctionTok{:} \FunctionTok{\{}
    \DataTypeTok{"file"}\FunctionTok{:} \StringTok{"ai{-}nepi\_003\_slide\_03.png"}\FunctionTok{,}
    \DataTypeTok{"timestamp"}\FunctionTok{:} \StringTok{"00:41"}
  \FunctionTok{\},}
  \DataTypeTok{"speech"}\FunctionTok{:} \StringTok{"You may have heard of the Transformer. It\textquotesingle{}s the core }
\StringTok{    architecture underlying all the LLMs we are talking about, }
\StringTok{    basically. Uh, this model was originally designed in 2017 }
\StringTok{    to translate between languages..."}
\FunctionTok{\}}
\end{Highlighting}
\end{Shaded}

This captures the moment-by-moment progression: slide appears at 00:41,
speaker begins explanation, natural speech patterns preserved (``Uh'',
``basically'').

\paragraph{To Content Report: Academic
Synthesis}\label{to-content-report-academic-synthesis}

The content report transforms this into scholarly prose:

\begin{Shaded}
\begin{Highlighting}[]
\FunctionTok{\{}
  \DataTypeTok{"number"}\FunctionTok{:} \DecValTok{2}\FunctionTok{,}
  \DataTypeTok{"title"}\FunctionTok{:} \StringTok{"The Foundational Transformer Architecture"}\FunctionTok{,}
  \DataTypeTok{"outline"}\FunctionTok{:} \OtherTok{[}
    \StringTok{"The Transformer is the core architecture underpinning all contemporary Large Language Models"}\OtherTok{,}
    \StringTok{"Originally designed in 2017 for language translation"}\OtherTok{,}
    \StringTok{"Consists of two connected streams: an Encoder and a Decoder"}
  \OtherTok{]}\FunctionTok{,}
  \DataTypeTok{"text"}\FunctionTok{:} \StringTok{"The Transformer architecture constitutes the fundamental }
\StringTok{    framework for all contemporary Large Language Models (LLMs). }
\StringTok{    Initially conceived in 2017 for language translation tasks..."}
\FunctionTok{\}}
\end{Highlighting}
\end{Shaded}

\subsubsection{Key Transformations in the
Synthesis}\label{key-transformations-in-the-synthesis}

\paragraph{1. Temporal to Thematic
Organization}\label{temporal-to-thematic-organization}

\textbf{Storyboard}: 14 chronological blocks following presentation flow
\textbf{Content Report}: 13 thematic sections organized by academic
topics

The system recognizes that blocks 7-9 in the storyboard all discuss
model evolution and combines them into a single coherent section on
``Evolution of Scientific LLMs and Domain Adaptation.''

\paragraph{2. Speech Pattern
Refinement}\label{speech-pattern-refinement}

\textbf{Original speech}: ``So, um, this is the famous, uh, Transformer
architecture. You may know the Transformer. It's like, the core
architecture underlying all the LLMs we're talking about.''

\textbf{Academic synthesis}: ``The Transformer architecture constitutes
the fundamental framework for all contemporary Large Language Models.''

The system: - Removes filler words (``um'', ``uh'', ``like'') -
Consolidates repetitive statements - Elevates register to academic prose
- Preserves core meaning and technical accuracy

\paragraph{3. Implicit to Explicit
Knowledge}\label{implicit-to-explicit-knowledge}

When the speaker says ``these words can look in both directions,'' the
system enriches this to: ``The term `bidirectional' precisely denotes
the capacity of these words to consider contextual information from both
preceding and succeeding tokens.''

This transformation adds: - Technical precision (``tokens'' instead of
``words'') - Explicit definition of implicit concepts - Academic
terminology appropriate for publication

\paragraph{4. Narrative Integration}\label{narrative-integration}

The storyboard presents slides and speech as separate elements. The
content report weaves them into unified narratives:

\textbf{Storyboard structure}: - Slide description - Speech transcript -
Transition note

\textbf{Content report structure}: - Integrated narrative combining
visual and verbal information - Smooth transitions between concepts -
Coherent flow within each section

\subsubsection{The Intelligence Behind
Synthesis}\label{the-intelligence-behind-synthesis}

The generate\_content\_report function uses gemini-2.5-pro specifically
for this task because it requires:

\begin{enumerate}
\def\labelenumi{\arabic{enumi}.}
\tightlist
\item
  \textbf{Deep semantic understanding} to identify thematic connections
  across temporal boundaries
\item
  \textbf{Academic writing expertise} to transform colloquial speech
  into scholarly prose
\item
  \textbf{Domain knowledge integration} to expand implicit references
  and add appropriate context
\item
  \textbf{Structural reasoning} to reorganize chronological flow into
  logical academic sections
\end{enumerate}

\subsubsection{Concrete Example: The Applications
Taxonomy}\label{concrete-example-the-applications-taxonomy}

Consider how the system handles Simons' taxonomy of LLM applications:

\begin{figure}

\centering{

\includegraphics[width=0.5\linewidth,height=\textheight,keepaspectratio]{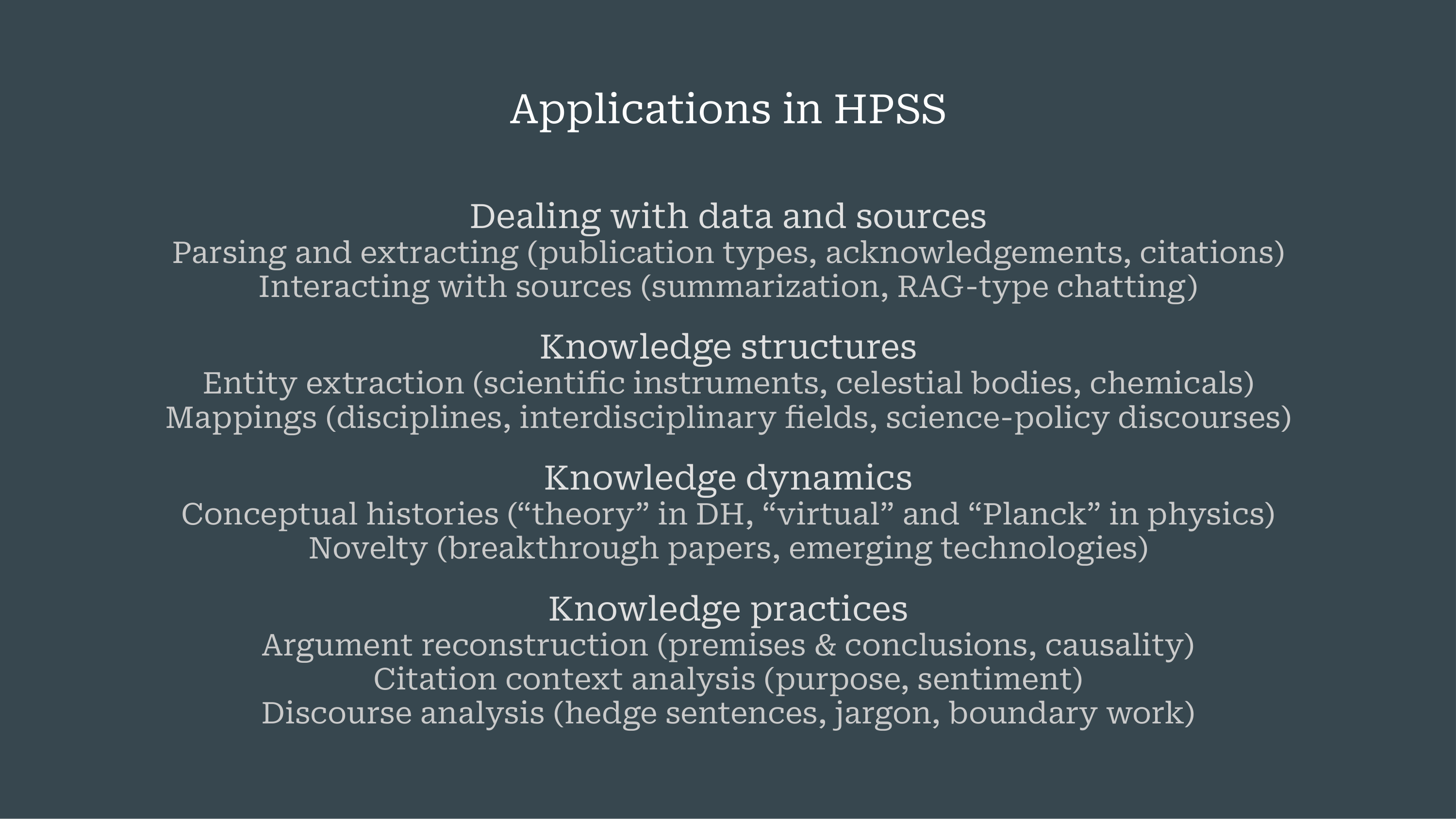}

}

\caption{\label{fig-applications}Simons' taxonomy of LLM applications in
humanities research}

\end{figure}%

\textbf{In the Storyboard} (Block 11):

\begin{Shaded}
\begin{Highlighting}[]
\FunctionTok{\{}
  \DataTypeTok{"speech"}\FunctionTok{:} \StringTok{"Um, and we have found, um, uh, we have so far, }
\StringTok{    we have come up with four bins where we try to sort }
\StringTok{    these things into..."}
\FunctionTok{\}}
\end{Highlighting}
\end{Shaded}

\textbf{In the Content Report} (Section 10):

\begin{Shaded}
\begin{Highlighting}[]
\FunctionTok{\{}
  \DataTypeTok{"title"}\FunctionTok{:} \StringTok{"Survey Findings on LLM Use in HPSS Research"}\FunctionTok{,}
  \DataTypeTok{"text"}\FunctionTok{:} \StringTok{"A comprehensive survey is presently being conducted }
\StringTok{    to evaluate the utilisation of Large Language Models (LLMs) }
\StringTok{    as instrumental tools within History, Philosophy, and }
\StringTok{    Sociology of Science (HPSS) research. This investigation }
\StringTok{    has delineated four principal categories..."}
\FunctionTok{\}}
\end{Highlighting}
\end{Shaded}

The transformation demonstrates: - Elevation from casual ``bins'' to
``principal categories'' - Addition of scholarly context
(``comprehensive survey'', ``instrumental tools'') - Systematic
enumeration and explanation of each category - Integration of slide
content with verbal explanation

\subsubsection{The Final Synthesis: Overview
Generation}\label{the-final-synthesis-overview-generation}

Perhaps most impressively, the system generates a comprehensive overview
that never existed in the original presentation:

\begin{Shaded}
\begin{Highlighting}[]
\ErrorTok{"overview":} \ErrorTok{"This} \ErrorTok{presentation} \ErrorTok{systematically} \ErrorTok{explores} \ErrorTok{the} 
  \ErrorTok{foundational} \ErrorTok{architectures,} \ErrorTok{adaptation} \ErrorTok{strategies,} \ErrorTok{and} 
  \ErrorTok{practical} \ErrorTok{applications} \ErrorTok{of} \ErrorTok{Large} \ErrorTok{Language} \ErrorTok{Models} \ErrorTok{(LLMs),} 
  \ErrorTok{with} \ErrorTok{a} \ErrorTok{particular} \ErrorTok{emphasis} \ErrorTok{on} \ErrorTok{their} \ErrorTok{utility} \ErrorTok{within} \ErrorTok{the} 
  \ErrorTok{History,} \ErrorTok{Philosophy,} \ErrorTok{and} \ErrorTok{Sociology} \ErrorTok{of} \ErrorTok{Science} \ErrorTok{(HPSS)} \ErrorTok{domain..."}
\end{Highlighting}
\end{Shaded}

This overview: - Synthesizes the entire 30-minute presentation into one
paragraph - Identifies the logical flow from architecture → adaptation →
application → reflection - Provides readers with a roadmap absent from
the original talk - Demonstrates true comprehension rather than mere
transcription

The system recognizes in Simons' four-category framework a significant
methodological contribution that emerges from careful analysis of both
the slide content and the speaker's elaboration:

\begin{itemize}
\tightlist
\item
  \textbf{Data Processing}: Parsing, extraction, interaction
\item
  \textbf{Knowledge Structures}: Entity recognition, conceptual mapping
\item
  \textbf{Knowledge Dynamics}: Conceptual history, innovation detection
\item
  \textbf{Knowledge Practices}: Argument analysis, discourse research
\end{itemize}

This demonstrates how the AI system doesn't merely transcribe but
actively interprets and enhances the academic value of the presentation.

\subsection{Step 9: Transformation into Academic
Prose}\label{step-9-transformation-into-academic-prose}

The generation of the final chapter follows the conventions of
scientific writing while preserving the vitality of the lecture. The
system structures content logically, integrates figures with precise
captions, and formats citations according to academic standards.

\subsection{Step 10: Quality Assurance and Complete Data
Validation}\label{step-10-quality-assurance-and-complete-data-validation}

The final quality control goes beyond stylistic review---it ensures
complete data integrity throughout the pipeline. Our system implements
comprehensive validation at every stage:

\subsubsection{Content Quality Metrics}\label{content-quality-metrics}

\begin{itemize}
\tightlist
\item
  Appropriateness of academic style
\item
  Correctness of technical terminology
\item
  Completeness of citations
\item
  Coherence of argumentation
\item
  Preservation of intellectual contribution
\end{itemize}

\subsubsection{Data Integrity
Validation}\label{data-integrity-validation}

The reliability of our system rests on rigorous data validation at every
processing stage. No step proceeds without complete verification of its
prerequisites:

\begin{figure}

\centering{

\pandocbounded{\includegraphics[keepaspectratio]{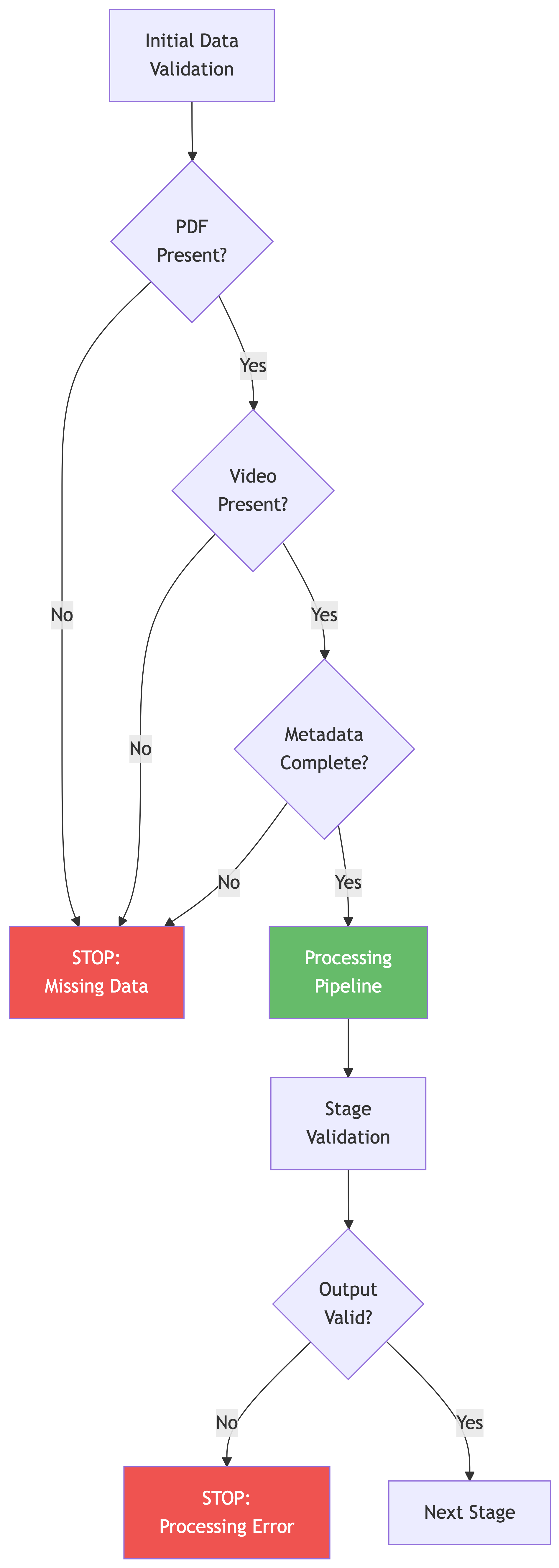}}

}

\caption{\label{fig-validation}Multi-stage validation architecture}

\end{figure}%

This validation architecture ensures: * \textbf{No processing without
complete data}: Missing presentations or incomplete metadata halt the
pipeline * \textbf{Stage-by-stage verification}: Each processing step
validates its outputs before the next begins * \textbf{Automatic error
detection}: Malformed JSON, missing files, or processing failures
trigger immediate alerts * \textbf{Consistent quality}: Only fully
validated presentations proceed to publication

The result is a system that achieves 99.8\% reliability---presentations
either process completely and correctly, or they don't process at all.
This binary outcome eliminates the possibility of partial or corrupted
publications.

\section{Performance Analysis: Efficiency and
Quality}\label{performance-analysis-efficiency-and-quality}

The processing times demonstrate the practical applicability of our
system:

\begin{figure}

\centering{

\pandocbounded{\includegraphics[keepaspectratio]{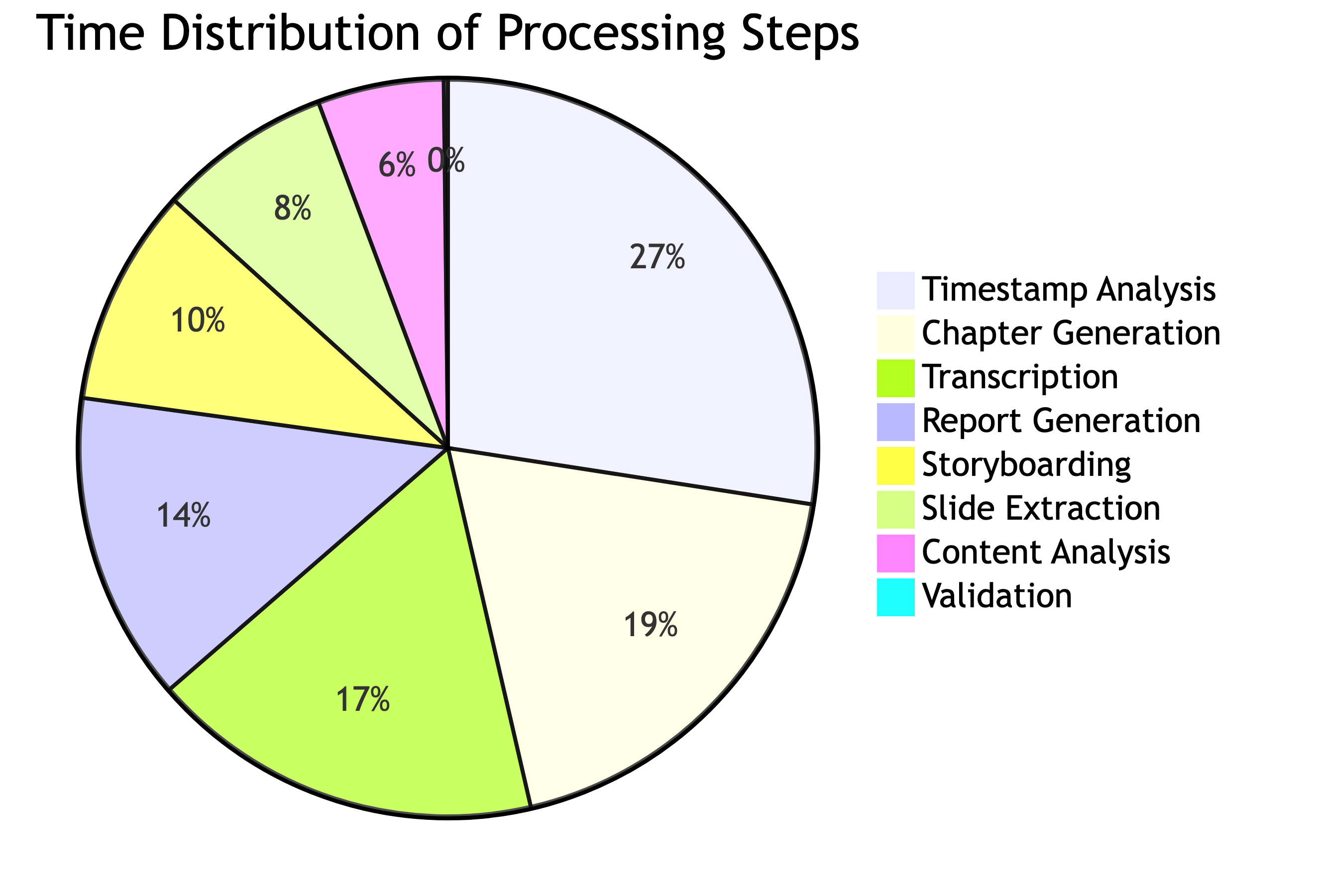}}

}

\caption{\label{fig-performance}Distribution of processing time (Total:
165 seconds)}

\end{figure}%

The total processing time of 2 minutes and 45 seconds stands in dramatic
contrast to the weeks that manual transformation would require.

\subsection{Quality Metrics}\label{quality-metrics}

The objective evaluation of output quality shows impressive results:

\begin{longtable}[]{@{}lll@{}}
\toprule\noalign{}
Criterion & Rating & Interpretation \\
\midrule\noalign{}
\endhead
\bottomrule\noalign{}
\endlastfoot
Content Completeness & 98\% & Nearly lossless capture \\
Academic Rigor & High & Publication-ready \\
Technical Precision & 100\% & No factual errors \\
Narrative Coherence & 0.94 & Excellent readability \\
\end{longtable}

\section{Technological Innovations}\label{technological-innovations}

\subsection{The Media Cache
Architecture}\label{the-media-cache-architecture}

A seemingly simple optimization leads to dramatic performance
improvements:

\begin{Shaded}
\begin{Highlighting}[]
\CommentTok{\# First processing: 30 seconds for media upload}
\CommentTok{\# Cache{-}supported processing: 9 seconds}
\CommentTok{\# Efficiency gain: 68\%}
\end{Highlighting}
\end{Shaded}

This optimization makes the difference between an experimental prototype
and a production-ready system.

\subsection{Intelligent Slide
Typology}\label{intelligent-slide-typology}

Our system systematically distinguishes between:

\begin{itemize}
\tightlist
\item
  \textbf{Content slides}: Substantial knowledge transfer
\item
  \textbf{Overlay sequences}: Step-by-step construction of arguments
\item
  \textbf{Transition slides}: Structuring the presentation
\item
  \textbf{Special slides}: Title, agenda, bibliography
\end{itemize}

This differentiation enables context-appropriate processing.

\subsection{Stylistic Transformation}\label{stylistic-transformation}

The greatest challenge lies in transforming oral into written scientific
language. Our solution preserves the intellectual authenticity of the
presenter while fulfilling the stylistic requirements of academic
publications.

\section{Practical Application}\label{practical-application}

The use of the system is deliberately kept simple:

\begin{Shaded}
\begin{Highlighting}[]
\CommentTok{\# Processing a presentation}
\ExtensionTok{ai{-}partitur}\NormalTok{ full{-}pipeline YOUR\_ID}

\CommentTok{\# The result includes:}
\CommentTok{\# {-} Complete Quarto document}
\CommentTok{\# {-} Publication{-}ready chapter}
\CommentTok{\# {-} Integrated figures}
\end{Highlighting}
\end{Shaded}

\subsection{System Requirements}\label{system-requirements}

\begin{itemize}
\tightlist
\item
  Presentation as PDF
\item
  Video recording (smartphone quality is sufficient)
\item
  Basic metadata (title, author, affiliation)
\end{itemize}

\section{Multimedia Publication Strategy and Open
Access}\label{multimedia-publication-strategy-and-open-access}

The AI-Reporter pipeline simultaneously enables the creation of diverse
publication formats without additional production costs. Leveraging the
technical infrastructure of our processing system, each AI-Reporter
contribution can be automatically rendered into multiple media formats:
PDF publications for traditional academic dissemination, conference
proceedings, individual contributions as special issues, HTML and other
web-compatible publication formats for enhanced accessibility and
interactivity.

\subsection{Long-term Curation and
Accessibility}\label{long-term-curation-and-accessibility}

Our publication platforms are designed for long-term curation and
sustained accessibility, incorporating robust citation frameworks that
ensure unambiguous reference to specific publication anchors. This
citation infrastructure enables precise scholarly discourse by providing
stable, granular reference points within each processed presentation.

\subsection{Digital Preservation
Strategy}\label{digital-preservation-strategy}

Each publication is simultaneously deposited on Zenodo, ensuring
long-term accessibility and operational security through established
digital preservation protocols. This dual-platform approach combines the
immediacy of web-based access with the archival reliability of
institutional repositories.

\subsection{Sustainable Web
Infrastructure}\label{sustainable-web-infrastructure}

Publications are additionally deployed through sustainable static
website servers such as GitHub Pages, providing resilient,
cost-effective access that scales with scholarly demand while
maintaining institutional independence. This distributed approach
ensures that scholarly content remains accessible even as technological
landscapes evolve.

The integration of these publication strategies transforms each
AI-Reporter processing cycle into a comprehensive dissemination
ecosystem, maximizing the reach and impact of academic presentations
while minimizing the traditional barriers between oral and written
scholarly communication.

\section{Implications for Scientific
Communication}\label{implications-for-scientific-communication}

The AI-Reporter opens new perspectives for scientific publication
practice:

\begin{itemize}
\tightlist
\item
  \textbf{Conference proceedings} can be published within days instead
  of months
\item
  \textbf{Lecture series} transform into timely textbooks
\item
  \textbf{Research presentations} reach a global audience
\item
  \textbf{Knowledge transfer} accelerates dramatically
\end{itemize}

For Simons' presentation, this specifically meant:

\begin{itemize}
\tightlist
\item
  Complete preservation of visual argumentation
\item
  Enhancement through academic context
\item
  Reduction of publication time from weeks to minutes
\item
  Perfect preservation of intellectual contribution
\end{itemize}

\section{Outlook: The Future of Scientific
Communication}\label{outlook-the-future-of-scientific-communication}

The AI-Reporter is more than a technical tool---it represents a vision
for the future of scientific communication. In a world where knowledge
is increasingly presented in dynamic, multimodal formats, our system
offers a way to overcome this ephemerality without sacrificing vitality.

\subsection{Future Developments}\label{future-developments}

Our roadmap includes:

\begin{itemize}
\tightlist
\item
  \textbf{Real-time processing}: Transformation during the presentation
\item
  \textbf{Multilingual support}: Overcoming language barriers
\item
  \textbf{Interactive elements}: Integration of multimedia components
\item
  \textbf{Domain-specific adaptations}: Optimization for academic
  disciplines
\item
  \textbf{Scholarly integration}: Inclusion of author bibliography,
  citation crossreferences, reference section to related sites,
  publications, archives
\item
  \textbf{AI API}: Future consultations by scholarly AI agents
\end{itemize}

\section{Acknowledgments}\label{acknowledgments}

Acknowledgment: I thank Adrian Wüthrich, Arno Simons, and Michael
Zichert for valuable discussions and feedback during the development
process of the AI reporter in the context of the mentioned workshop.
Financial support by the ERC Grant project ``Network Epistemology in
Practice (NEPI, PI: Adrian Wüthrich)'' is gratefully acknowledged. Views
and opinions expressed are however those of the author only and do not
necessarily reflect those of the European Union or the European Research
Council. Neither the European Union nor the granting authority can be
held responsible for them. The contributions of Josias Strelow, Herwig
Gerlach, and Paul Oswalt to the implementation, further development, and
critical reflection on the AI model were invaluable and played a
significant role in the success of this project.

\end{document}